# Extended gamma-ray sources around pulsars constrain the origin of the positron flux at Earth

**Authors:** A.U. Abeysekara[a], A. Albert[b], R. Alfaro[c], C. Alvarez[d], J.D. Álvarez[e], R. Arceo[d], J.C. Arteaga-Velázquez[e], D. Avila Rojas[c], H.A. Ayala Solares[f], A.S. Barber[a], N. Bautista-Elivar[g], A. Becerril[c], E. Belmont-Moreno[c], S.Y. BenZvi[h], D. Berley[i], A. Bernal[j], J. Braun[k], C. Brisbois[f], K.S. Caballero-Mora[d], T. Capistrán[l], A. Carramiñana[l], S. Casanova[m, n], M. Castillo[e], U. Cotti[e], J. Cotzomi[o], S. Coutiño de León[l], C. De León[o], E. De la Fuente[p], B.L. Dingus[b], M.A. DuVernois[k], J.C. Díaz-Vélez[p], R.W. Ellsworth[q], K. Engel[i], O. Enríquez-Rivera[r], D.W. Fiorino[i], N. Fraija[j], J.A. García-González[c], F. Garfias[j], M. Gerhardt[f], A. González Muñoz[c], M.M. González[j], J.A. Goodman[i], Z. Hampel-Arias[k], J.P. Harding[b], S. Hernández[c], A. Hernández-Almada[c], J. Hinton[n], B. Hona[f], C.M. Hui[s], P. Hüntemeyer[f], A. Iriarte[j], A. Jardin-Blicq[n], V. Joshi[n], S. Kaufmann[d], D. Kieda[a], A. Lara[r], R.J. Lauer[t], W.H. Lee[j], D. Lennarz[u], H. León Vargas[c], J.T. Linnemann[v], A.L. Longinotti[l], G. Luis Raya[g], R. Luna-García[w], R. López-Coto[n] *, K. Malone[x], S.S. Marinelli[v], O. Martinez[o], I. Martinez-Castellanos[i], J. Martínez-Castro[w], H. Martínez-Huerta[y], J.A. Matthews[t], P. Miranda-Romagnoli[z], E. Moreno[o], M. Mostafá[x], L. Nellen[aa], M. Newbold[a], M.U. Nisa[h], R. Noriega-Papaqui[z], R. Pelayo[w], J. Pretz[x], E.G. Pérez-Pérez[g], Z. Ren[t], C.D. Rho[h], C. Rivière[i], D. Rosa-González[l], M. Rosenberg[x], E. Ruiz-Velasco[c], H. Salazar[o], F. Salesa Greus[m] *, A. Sandoval[c], M. Schneider[ab], H. Schoorlemmer[n], G. Sinnis[b], A.J. Smith[i], R.W. Springer[a], P. Surajbali[n], I. Taboada[u], O. Tibolla[d], K. Tollefson[v], I. Torres[l], T.N. Ukwatta[b], G. Vianello[ac], T. Weisgarber[k], S. Westerhoff[k], I. G. Wisher[k], J. Wood[k], T. Yapici[v], G. Yodh[ad], P.W. Younk[b], A. Zepeda[y,d] , H. Zhou[b]*, F. Guo[b], J. Hahn[n], H. Li[b], H. Zhang[b]

[a]Department of Physics and Astronomy, University of Utah, Salt Lake City, UT, USA
[b]Los Alamos National Laboratory, Los Alamos, NM, USA
[c]Instituto de Física, Universidad Nacional Autónoma de México, Mexico City, Mexico
[d]Universidad Autónoma de Chiapas, Tuxtla Gutiérrez, Chiapas, Mexico
[e]Universidad Michoacana de San Nicolás de Hidalgo, Morelia, Mexico
[f]Department of Physics, Michigan Technological University, Houghton, MI, USA
[g]Universidad Politecnica de Pachuca, Pachuca, Hidalgo, Mexico
[h]Department of Physics & Astronomy, University of Rochester, Rochester, NY, USA
[i]Department of Physics, University of Maryland, College Park, MD, USA
[j]Instituto de Astronomía, Universidad Nacional Autónoma de México, Mexico City, Mexico
[k]Department of Physics, University of Wisconsin-Madison, Madison, WI, USA
[l]Instituto Nacional de Astrofísica, Óptica y Electrónica, Puebla, Mexico
[m]Institute of Nuclear Physics Polish Academy of Sciences, Krakow, Poland
[n]Max-Planck Institute for Nuclear Physics, Heidelberg, Germany
[o]Facultad de Ciencias Físico Matemáticas, Benemérita Universidad Autónoma de Puebla, Puebla, Mexico
[p]Departamento de Física, Centro Universitario de Ciencias Exactas e Ingenierías, Universidad de Guadalajara, Guadalajara, Mexico
[q]School of Physics, Astronomy, and Computational Sciences, George Mason University, Fairfax, VA, USA
[r]Instituto de Geofísica, Universidad Nacional Autónoma de México, Mexico City, Mexico
[s]Astrophysics Office, NASA Marshall Space Flight Center, Huntsville, AL, USA
[t]Department of Physics and Astronomy, University of New Mexico, Albuquerque, NM, USA
[u]School of Physics and Center for Relativistic Astrophysics - Georgia Institute of

Technology, Atlanta, GA, USA
[v]Department of Physics and Astronomy, Michigan State University, East Lansing, MI, USA
[w]Centro de Investigación en Computación, Instituto Politécnico Nacional, Mexico City, Mexico.
[x]Department of Physics, Pennsylvania State University, University Park, PA, USA
[y]Physics Department, Centro de Investigacion y de Estudios Avanzados del IPN, Mexico City, Mexico
[z]Universidad Autónoma del Estado de Hidalgo, Pachuca, Mexico
[aa]Instituto de Ciencias Nucleares, Universidad Nacional Autónoma de México, Mexico City, Mexico
[ab]Santa Cruz Institute for Particle Physics, University of California, Santa Cruz, Santa Cruz, CA, USA
[ac]Hansen Experimental Physics Laboratory, Stanford University, Stanford, CA, USA
[ad]Department of Physics and Astronomy, University of California, Irvine, CA, USA

* Correspondence to: francisco.salesa@ifj.edu.pl, rlopez@mpi-hd.mpg.de, hao@lanl.gov

**Abstract:**
**The unexpectedly high flux of cosmic ray positrons detected at Earth may originate from nearby astrophysical sources, dark matter, or unknown processes of cosmic-ray secondary production. We report the detection, using the High-Altitude Water Cherenkov Observatory (HAWC), of extended tera-electron volt gamma-ray emission coincident with the locations of two nearby middle-aged pulsars (Geminga and PSR B0656+14). The HAWC observations demonstrate that these pulsars are indeed local sources of accelerated leptons, but the measured tera-electron volt emission profile constrains the diffusion of particles away from these sources to be much slower than previously assumed. We demonstrate that the leptons emitted by these objects are therefore unlikely to be the origin of the excess positrons, which may have a more exotic origin.**

**One Sentence summary:**
HAWC detection of extended gamma-ray emission from two very nearby middle-aged energetic pulsars disfavors them as origin of the positron excess above GeV energies detected at the Earth.

**Main Text:**
Cosmic rays are high-energy particles from space that have been known for more than a century. The origin of high-energy cosmic rays and how they are accelerated remains unclear. Most cosmic rays are protons or atomic nuclei, but positrons and electrons also are a small fraction of the total cosmic ray flux. Positrons are especially puzzling because the PAMELA (Payload for Antimatter Exploration and Light-nuclei Astrophysics) detector observed an unexpected excess of positrons at energies >10 GeV, compared with the predicted flux that originates from interactions of cosmic-ray protons propagating through the Galaxy (1). Confirmation of these results has come from the Fermi Large Area Telescope (2) and AMS [Alpha Magnetic Spectrometer (3)] experiments; the latter also showed that the excess signal extends to hundreds of giga-electron volts.

Energy losses experienced in interstellar magnetic and radiation fields by the highest-energy positrons require that their sources lie within a few hundred parsecs from Earth (4). Nearby potential cosmic ray accelerators for example, pulsar wind nebulae (PWNe) have been proposed as the sources of these extra positrons (5; 6). A PWN consist of a rapidly spinning neutron star (pulsar) that produces a wind of electrons and positrons that are further accelerated by the surrounding shock with the interstellar medium (ISM). There are a handful of known pulsars that are both close enough to be candidate sources and sufficiently old for the highest energy positrons to have had time to arrive at Earth (7; 8). Nearby dark matter particle interactions could also produce positrons (9). Both PWNe and dark matter sources should also produce gamma rays that could potentially be observed coming from the sources, unlike positrons (whose paths are deflected by magnetic fields).

Recently, the High-Altitude Water Cherenkov Observatory (HAWC) collaboration reported the detection of tera-electron volt gamma rays around two nearby pulsars, which are among those proposed to produce the local positrons (10). HAWC is a wide field-of-view, continuously operating detector of extensive air showers initiated by gamma rays and cosmic rays interacting in the atmosphere (11). The angular resolution improves from 1.0° to 0.2° with the size of the air shower. HAWC is the most sensitive survey detector above 10 TeV and is well suited to detecting nearby sources, which would have a greater angular extent. Operation of the full detector began in March 2015, and the data set presented here includes 507 days, as described in (11).

Tera-electron volt gamma-ray emissions from the pulsars Geminga and PSR B0656+14 were found in a search for extended sources that was performed for the HAWC catalog, in which these two pulsars have the designations 2HWC J0635+180 and 2HWC J0700+143 (10). By fitting to a diffusion model (12), the two sources were detected with a significance at the pulsar location of 13.1 and 8.1 standard deviations (σ), respectively (Fig. 1A). The tera-electron volt emission region is several degrees across which we attribute to electrons and positrons diffusing away from the pulsar and upscattering the cosmic microwave background (CMB) photons. Geminga was previously detected at tera-electron volt energies by the Milagro observatory, with a flux and angular extent consistent with the HAWC observation but with lower statistical significance (13). Here we show that the HAWC observation of the spectral and spatial properties of these sources can be used to constrain their contribution to the positron flux at Earth (Fig. 1B).

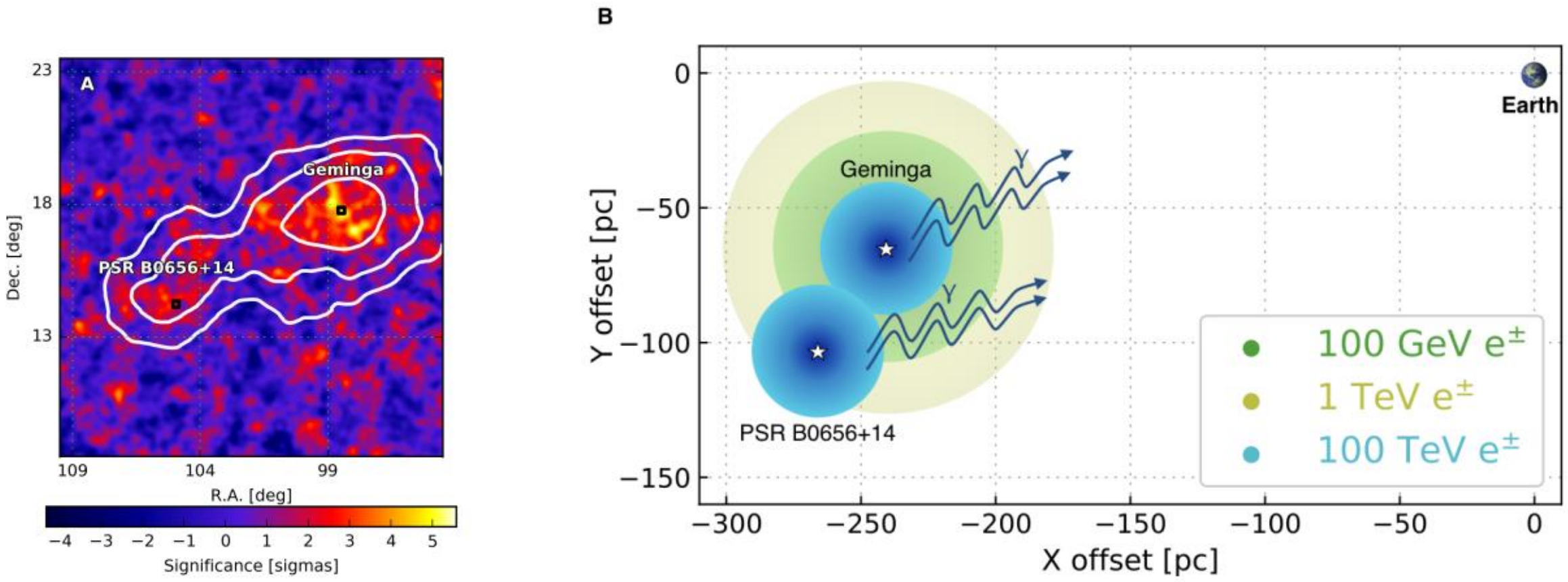

Fig. 1: Spatial Morphology of Geminga and PSR B0656+14. (A) HAWC significance map (between 1 and 50 TeV) for the region around Geminga and PSR B0656+14, convolved with the HAWC point spread function and with contours of 5σ, 7σ, and 10σ for a fit to the diffusion model. R.A., right ascension; dec., declination (B) Schematic illustration of the observed region and Earth, shown projected onto the Galactic plane. The colored circles correspond to the diffusion distance of leptons with three different energies from Geminga; for clarity, only the highest energy (blue) is shown for PSR B0656+14. The balance between diffusion rate and cooling effects means that tera-electron volt particles diffuse the farthest (Fig. S1).

| **Pulsar Parameters** | | **Geminga** | **PSR B0656+14** |
|---|---|---|---|
| (Right ascension, declination) (J2000 source location) | [degrees] | (98.48, 17.77) | (104.95, 14.24) |
| $\tau_c$ (characteristic age) | [years] | 342,000 | 110,000 |
| T (spin period) | [seconds] | 0.237 | 0.385 |
| d (distance) | [parsecs] | $250^{+120}_{-62}$ | $288^{+33}_{-27}$ |
| dE/dt (energy loss rate due to pulsar's spin slowing) | [x$10^{34}$ ergs/sec] | 3.26 | 3.8 |
| **Model Values** | | | |
| $\theta_0$ ($\theta_d$ for 20 TeV $\gamma$-ray) | [degrees] | 5.5 ± 0.7 | 4.8 ± 0.6 |
| $N_0$ | [x$10^{-15}$ photons/TeV/cm$^2$ /sec] | $13.6^{+2.0}_{-1.7}$ | $5.6^{+2.5}_{-1.7}$ |
| $\alpha$ | | 2.34 ± 0.07 | 2.14 ± 0.23 |
| $D_{100}$ (Diffusion coefficient of 100TeV electrons from joint fit of two PWNe) | [x$10^{27}$ cm$^2$/sec] | 4.5 ± 1.2 | 4.5 ± 1.2 |
| $D_{100}$(Diffusion coefficient of 100TeV electrons from individual fit of PWN) | [x$10^{27}$ cm$^2$/sec] | $3.2^{+1.4}_{-1.0}$ | $15^{+49}_{-9}$ |
| Energy Range | [TeV] | 8 to 40 | 8 to 40 |
| Luminosity in gamma-rays over this energy range | [x$10^{31}$ erg/sec] | 11x(d/250 parsec)$^2$ | 4.5x(d/288 parsec) $^2$ |
| **Assumed Parameters** | | | |
| $L_0$ (initial spin down power) | [x$10^{36}$ ergs/sec] | 27.8 | 4.0 |
| $W_e$ (total energy released since pulsar's birth) | [x$10^{48}$ ergs] | 11.0 | 1.5 |

Table 1: Pulsar parameters, values of parameters from the model fitting to the observed extended gamma-ray emission, and assumed parameters of our model. Pulsar parameters are from (14).

A diffusion model of the spatial and spectral morphology (see below and (12)) is fit to the gamma-ray flux *N* as a function of angle $\theta$ from the source and gamma-ray energy *E* as

$$\frac{d^2N}{dEd\Omega} = N_o\left(\frac{E}{20\text{ TeV}}\right)^{-\alpha}\frac{1.22}{\pi^{3/2}\theta_d(E)(\theta+0.06\theta_d(E))}\exp(-\theta^2/\theta_d(E)^2)\ (1)$$

using a maximum likelihood technique. $N_0$ is the flux normalization at 20TeV, and Ω denotes a solid angle. The diffusion angle $\theta_d$ is proportional to the square root of the diffusion coefficient *D,* and both vary with energy. The model values from the fit are given in Table 1. The spectral indices $\alpha$ and observed fluxes are similar to those of other tera-electron volt PWNe (15), but the luminosities are lower, primarily because of their nearby distance and larger apparent size. The energy range is estimated by increasing (decreasing) the minimum (maximum) energy of an abrupt cut-off in the power law spectrum until the significance of the fit decreases by 1σ.

Assuming that all the observed gamma-ray emission at tera-electron volt energies is produced by relativistic electron and positron pairs, we calculate the electron and positron flux produced by these sources at Earth. Tera-electron volt gamma rays are produced when positrons or electrons inverse Compton scatter lower-energy photons to higher energies. Gamma rays at ~20 TeV are produced by electrons and positrons at ~100 TeV (12). At these energies, the scattered photons are primarily from the CMB, because the cross-section for scattering higher energy infrared and optical photons is strongly suppressed.

Charged particles also lose energy by synchrotron radiation when propagating through a magnetic field. At the nominal distances of Geminga and PSR B0656+14 at 250 and 288 pc, respectively (14), the observed spatial extent of these two sources at TeV energies is tens of parsecs which is much greater than the <0.1 pc nebula observed in x-rays (16; 17). The x-ray emission is from synchrotron radiation where the magnetic field is enhanced by the pulsar to 10-20 μG (18). The region emitting tera-electron volt gamma rays is primarily outside the x-ray nebula, so we assume that the magnetic field is equal to that of the nearby ISM at 3 μG (19) and is not increased by the presence of the pulsar or the prior supernova remnant. The implied energy density of the particles accelerated by the pulsar is several orders of magnitude less than the energy density of the interstellar magnetic fields, so the particles themselves do not amplify the field. If the magnetic field was higher, the pulsar could not provide enough energy to produce the observed gamma-ray luminosity.

We assume that particles are accelerated from 1 GeV to 500 TeV with a power-law energy spectrum. The spectral index of the electrons is chosen to fit the HAWC gamma-ray observations and is harder than the gamma-ray spectra by ~0.1 (12). The total flux of electrons and positrons is 40 and 4% of the measured power released by the slowing of the spin of Geminga and PSR B0656+14, respectively. Because most of the power released is in electrons with energies below those measured by HAWC, the conversion of the spin-down power to accelerated electrons could be less efficient if the spectrum breaks at lower energies. However, lower efficiencies reduce the local positron flux. The tera-electron volt gamma rays could also be produced by high-energy protons interacting with matter. However, the matter density around these sources is low (20), and a proton origin of the gamma-ray emission is disfavored because the required power would be even larger than the total energy injected by the pulsar.

If electrons and positrons are injected continuously and diffuse isotropically away from their sources, the density of particles is proportional to $\frac{1}{r}\,\mathrm{erfc}(r/r_d)$, where $r_d$ is the diffusion radius, r is the distance from the pulsar, and erfc is the error function (21). The integral of this function along the line of sight is the gamma-ray surface brightness. The diffusion radius is defined as $r_d = 2\sqrt{D\,t}$; here, $D$ is the diffusion coefficient, and $t$ is the lepton injection time and is chosen to be the electron-cooling time for the energy of electrons that produce the HAWC observation (12). The diffusion coefficient increases with energy $E$ as $E^{\delta}$, where $\delta$ is chosen to be 1/3, motivated by the Kolmogorov turbulence model (22). This value of $\delta$ is also compatible with recent results from AMS-02 for the spectrum of hadronic cosmic rays that are produced by spallation (23).

The observed surface brightness distributions of Geminga and PSR B0656+14 are shown in Fig. 2, along with our model fit. The fitted values of $D_{100}$ (where $D_{100}$ is the diffusion coefficient for electrons at 100 TeV) are given in Table 1. Because the sources are close to each other spatially, as seen in Fig. 1, we also fit a single $D_{100}$ to both (Table 1). The values of $D_{100}$ derived from a joint fit and fits to the individual sources are equivalent within the uncertainties. We also fit the emission around Geminga with an elongated diffusion model and the fit result is not statistically significant over the isotropic diffusion model, so there is no evidence of anisotropic diffusion in these data.

The value of $D_{100}$ derived from our HAWC observations ($4.5\pm1.2 \times 10^{27}$ cm$^2$ s$^{-1}$) is smaller by a factor of about 100 than those considered in previous models of electron diffusion into the local ISM (5; 6; 7; 8; 24). These other models assumed that D was similar to the value inferred from hadronic cosmic rays, which may not be applicable to positrons in the local ISM. Spatial inhomogeneities are possible (25), and such a low D could arise from additional effects of turbulent scattering (26; 27), for example. Because the angular extent of the TeV source is proportional to $\sqrt{D_{100}}$, a diffusion coefficient larger by a factor of 100 would result in an angular extent for the source that is larger by a factor of 10, and a surface brightness for the same total flux that is smaller by a factor of 100. This would make these two sources undetectable by HAWC.

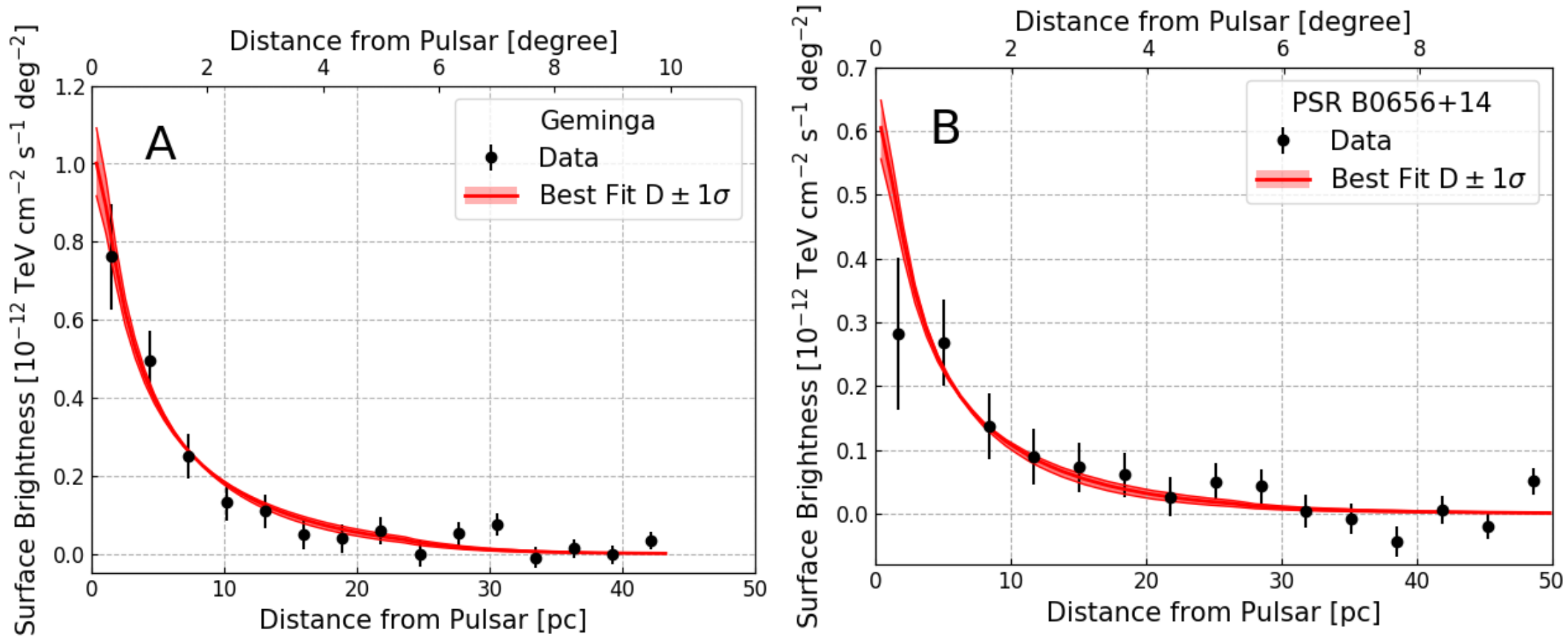

Fig. 2: Surface brightness of the tera-electron volt gamma-ray emission. Surface brightness is shown as a function of the distance from the Geminga pulsar (A) and PSR B0656+14 (B). The solid line represents the best fitting model with a common diffusion coefficient, and the shaded band is the ±1σ statistical uncertainty. Error bars are statistical. The distance from each pulsar in parsecs is calculated based on nominal distances of 250 and 288 pc for Geminga and PSR B0656+14, respectively (14).

To calculate the positrons that have diffused to Earth, the history of the pulsar's emission must be included because the lifetime of sub-tera-electron volt positrons in the ISM can exceed that of the pulsar. Assuming that a pulsar is a pure dipole radiator and hence has a braking index of 3, its luminosity $L$ at a time $t$ after its birth is predicted to vary as $L=L_0(1+t/\tau)^{-2}$. We take the characteristic initial pulsar spin-down timescale ($\tau$) of 12,000 years for Geminga (28) and assume it to be the same for PSR B0656+14. The electron transport equation is solved using the EDGE code (29) for electron diffusion (12).

Figure 3 shows the expected flux of positrons as a function of energy from Geminga (blue line) compared with the measured flux of positrons by AMS-02 in low Earth orbit. The positron flux from Geminga exceeds by several orders of magnitude that from PSR B0656+14, owing to the combination of Geminga's greater gamma-ray flux that injects more energy into electrons, its older age and its closer distance. We consider the impact of different systemic effects (12): if the spectral index of the diffusion coefficient $\delta$ were smaller, lower energy positrons would diffuse faster; if the characteristic initial spin-down timescale $\tau$ were shorter, the luminosity would have been higher in the past. If the current distance were smaller that would not change the local positron flux substantially because the true $D_{100}$ would also have to be smaller (because it is derived from the angular extent of the sources). Therefore, in this model, these pulsars do not produce a measurable contribution to the positron flux measured by AMS-02 at Earth. Moreover, regardless of the absolute flux, their strongly peaked energy spectrum is incompatible with the flat distribution in energy found by AMS-02.

Our conclusions conflict with a recent estimate (30) based on the HAWC catalog data (*10*). The model in (30) considers the effect of constant-velocity convective winds to be dominant over diffusion. We have shown here that the tera-electron volt gamma-ray surface brightness measured by HAWC not available at the time of publication of (30) is well fit by a diffusion profile and strongly disfavors the convection profile. Moreover, the absence of any pressure or energy source to power this strong wind strongly disfavors the model in (30) and the conclusions drawn regarding the local positron flux.

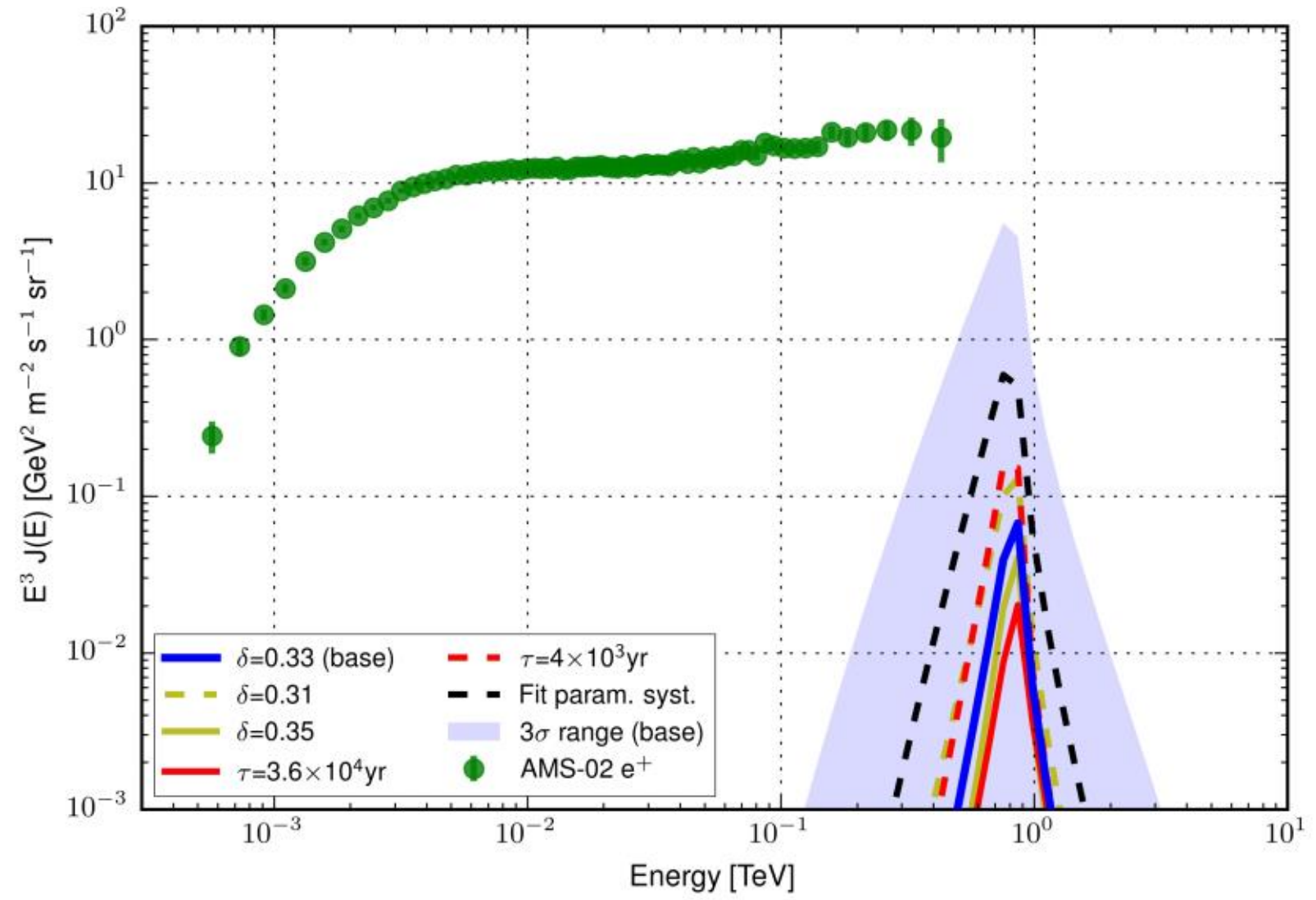

Fig. 3: Estimated positron energy flux at Earth from Geminga (blue solid line), compared with AMS-02 experimental measurements (green dots). The shaded blue region indicates the 3σ (99.5 % confidence) statistical uncertainty from simulations (12). Additional lines represent the effect on the positron energy flux of varying several systematic effects: the pulsar characteristic initial spin-down timescale (red lines); the index of the diffusion coefficient (yellow lines); and the diffusion angle, spectral index of the injected electrons, and flux normalization combined (black line). The spectrum expected Earth is strongly peaked at the energy where the cooling time equals the age of the pulsar. The positron flux at the Earth from PSR B0656+14 is several orders of magnitude lower than that from Geminga and below the range of this plot. AMS-02 experimental measurements are from (31); error bars come from the quadratic sum of statistical and systematic errors. Both AMS-02 experimental measurements and error values were fetched from an online database (32).

Geminga and PSR B0656+14 are the oldest pulsars for which a tera-electron volt nebula has so far been detected. Under our assumption of isotropic and homogeneous diffusion, the dominant source of the positron flux above 10 GeV cannot be either Geminga or PSR B0656+14. Under the unlikely situation that the field is nearly aligned along the direction between Earth and the nearby tera-electron volt nebulae, the local positron flux can be increased; however, the tera-electron volt morphology of the sources matches our isotropic diffusion model. We therefore favor the explanation that instead of these two pulsars, the origin of the local positron flux must be explained by other processes, such as different assumptions about secondary production [although that has been questioned (33; 34)], other pulsars, other types of cosmic accelerators such as micro-quasars (35) and supernova remnants (34), or the annihilation or decay of dark matter particles (9).

**Supplementary Materials**
www.sciencemag.org
Materials and Methods
Figs. S1, S2
References (36-41) [Note: The numbers refer to any additional references cited only within the Supplementary Materials]

**Acknowledgements:** We acknowledge the support from: the US National Science Foundation (NSF); the US Department of Energy Office of High-Energy Physics; the Laboratory Directed Research and Development program of Los Alamos National Laboratory; Consejo Nacional de Ciencia y Tecnología, México (grants 271051, 232656, 260378, 179588, 239762, 254964, 271737, 258865, 243290, 132197, and 281653) (Cátedras 873, 1563); Laboratorio Nacional HAWC de rayos gamma; L'OREAL Fellowship for Women in Science 2014; Red HAWC, México; DGAPA-UNAM (Dirección General Asuntos del Personal Académico-Universidad Nacional Autónoma de México; grants IG100317, IN111315, IN111716-3, IA102715, 109916, IA102917); VIEP-BUAP (Vicerrectoría de Investigación y Estudios de Posgrado-Benemérita Universidad Autónoma de Puebla); PIFI (Programa Integral de Fortalecimiento Institucional) 2012 and 2013; PROFOCIE (Programa de Fortalecimiento de la Calidad en Instituciones Educativas) 2014 and 2015; the University of Wisconsin Alumni Research Foundation; the Institute of Geophysics, Planetary Physics, and Signatures at Los Alamos National Laboratory; Polish Science Centre grant DEC-2014/13/B/ST9/945; and Coordinación de la Investigación Científica de la Universidad Michoacana. Thanks to S. Delay, L. Díaz, and E. Murrieta for technical support.
The data and software to reproduce Figs. 1 to 3 and the values of the gamma-ray flux parameters in Table 1 are available at https://data.hawc-observatory.org/datasets/geminga2017/

## Materials and Methods:

**Likelihood analysis**: The HAWC analysis framework (*36*) is based on a maximum likelihood model used for both the flux estimation and the morphology studies. The three pieces of information used in the analysis are: observed data and background estimate (using a HEALPix map (*37*)), the physical model (e.g. a power law with a given morphology), and the detector response, which parameterizes the energy resolution and point spread function from Monte Carlo simulations.
The likelihood L is computed as:

$$\ln L(\varphi; N_{\mathrm{obs}}) = \sum_{B=1}^{9} \sum_{p=1}^{p=\mathrm{last}} \ln\left(P\left((N_{\mathrm{obs}})_p|\varphi\right)\right), \quad \text{(S1)}$$

where *B* is the analysis bin number (*11*), *φ* are the flux normalization, spectral index, and diffusion angle parameters in our model ($N_0^{\mathrm{Gem}}$, $\alpha^{\mathrm{Gem}}$, $N_0^{\mathrm{PSR}}$, $\alpha^{\mathrm{PSR}}$, and $\theta_0$ where the superscript refers to Geminga or PSR B0656+14), $(N_{\mathrm{obs}})_p$ is the number of observed events in each pixel *p*, $N_{\mathrm{obs}}$ is the total number of events, and *P* is the Poisson probability of detecting $(N_{\mathrm{obs}})_p$ for a given set of parameters *φ*. Once the parameters that maximize the value of ln*(L)* are computed, the model used (the alternative hypothesis) is compared to a background-only model (the null hypothesis) using the Likelihood Ratio test. Our test statistic is defined as:

$$TS = 2\ln\frac{L(\mathrm{AlternativeHypothesis}; N_{\mathrm{obs}})}{L(\mathrm{NullHypothesis}; N_{\mathrm{obs}})} \quad \text{(S2).}$$

and is used to compute the significances in Fig.1.

**Electron diffusion model**: The gamma-ray spectrum and morphology are determined from a model of $e^+e^-$ pairs diffusing into the interstellar medium around the pulsar. The leptons produce gamma rays through inverse Compton scattering of CMB, infrared (IR), and optical photons. The photon fields are considered to have a gray body distribution with temperature (*T*) and energy density ($w_e$) and the values are approximately those derived by GALPROP (*4*):

| | T(K) | $w_e$ (eV/cm$^3$) |
|---|---|---|
| CMB | 2.7 | 0.26 |
| IR | 20 | 0.3 |
| Optical | 5000 | 0.3 |

Table S1: Values of the photon field parameters assumed in our model.

In the case of continuous injection of electrons and positrons from a point source at a constant rate

$dN_e/dE=Q_0E_e^{-\Gamma}$ (S3)

where $N_e$ is the number of electrons with an energy *E*, $Q_0$ is the normalization of the flux, and *Γ* is the spectral index. The radial distribution of the electrons at an instant *t* and distance *r* from the source depends on the energy of the electrons and positrons $E_e$ ((*21*), their equation 21):

$f(t,r,E_{\rm e}) = \frac{Q_0 E_{\rm e}^{-\Gamma}}{4\pi D(E_{\rm e}) r} erfc(r/r_d)$ (S4)

where $D(E_e)$ is the diffusion coefficient and $r_d$ the diffusion radius, defined as:

$D(E_{\rm e}) = D_0 (E_{\rm e}/10\ GeV)^{\delta}$ (S5).
$r_d = 2\sqrt{D(E_{\rm e}) t_E}$ (S6).

The spectral index of the diffusion coefficient is chosen to be $\delta$ = 1/3, motivated by the Kolmogorov turbulence model (*22*). $t_E$ is the smaller of two timescales: the injection time $t$ (in this case the age of the pulsar), or the cooling time $t_{\rm cool}$ (the lifetime of the electrons in the interstellar medium after synchrotron and inverse Compton losses). The cooling time, when Klein-Nishina effects are important (*37*), is:
$t_{\rm cool} = \frac{m_{\rm e} c^2}{4/3 c \sigma_{\rm T} \gamma} \cdot \frac{1}{u_B + u_{\rm ph}/(1+4\gamma\varepsilon_o)^{3/2}}$ (S7)

where $\sigma_{\rm T}$ is the Thomson cross section, $\gamma$ is the bulk Lorentz factor of the electron, $m_e$ is the mass of the electron, c is the speed of light, $u_B = B^2/8\pi$ is the magnetic density, $u_{ph}$ is the CMB energy density and $\varepsilon_0$ is the normalized CMB photon energy. The Inverse Compton losses from IR and optical photons are highly suppressed by Klein-Nishina effects and therefore negligible. Electrons of 100 TeV have a cooling time (~10 kyr) which is smaller than the age of the pulsar, so $t_E = t_{\rm cool}$.

The diffusion radius as a function of the energy for Geminga is shown in Fig. S1. Uncooled electrons increase their diffusion radius with energy up to the energy at which the cooling time equals the age of the pulsar; thereafter the diffusion radius decreases with energy.

TeV gamma-Ray morphology: Integrating the energy distribution in Eq. S3 along the observer line-of-sight from the electron source to Earth, the gamma-ray flux $f_d$ as a function of the distance $d$ is approximately proportional to:

$$f_d = \frac{1.22}{\pi^{3/2} r_d (d + 0.06 r_d)} exp(-d^2/r_d^2) \quad \text{(S8)}$$

We define the diffusion angle $\theta_d$ from the diffusion radius as

$$\theta_d \equiv \frac{180^o}{\pi} \cdot \frac{r_d}{d_{\rm src}} \quad \text{(S9)}$$

with $d_{src}$ the distance from the source to Earth. Equation S8 becomes:
$$f_\theta = \frac{1.22}{\pi^{3/2} \theta_d (\theta + 0.06\theta_d)} exp(-\theta^2/\theta_d^2) \quad \text{(S10)},$$

where $\theta$ is the angular distance from the pulsar. The diffusion angle becomes:

$$\theta_d = \theta_0 \left(\frac{E_{\rm e}}{E_{\rm e0}}\right)^{\frac{\delta-1}{2}} \sqrt{\frac{B^2/8\pi + u_{\rm ph}/(1+4\varepsilon_0 E_{\rm e0}/m_{\rm e}c^2)^{3/2}}{B^2/8\pi + u_{\rm ph}/(1+4\varepsilon_0 E_e/m_{\rm e}c^2)^{3/2}}} \quad \text{(S11)}$$

where $\theta_0$ is the diffusion angle at a pivot electron energy of $E_{e0}$.The relation between the mean electron and gamma-ray energy in the inverse Compton scattering process is given by (*39*):

$$< E_e > \approx 17 < E_\gamma >^{0.54+0.046\log_{10}(<E_\gamma>/TeV)} \quad \text{(S12)}$$

The pivot electron energy $E_{e0}$ is arbitrary but is chosen at 100 TeV, which corresponds to a gamma-ray energy of $E_0$~20 TeV, because it corresponds to the typical gamma-ray energy detected by HAWC.

The gamma-ray emission of Geminga and PSR B0656+14 are fit using 3ML (Multi-Mission Maximum Likelihood) (*40*) to the model morphology $f_\theta$ in Eq. S9 and S10 with a simple power law energy spectrum, using the average ISM magnetic field of 3μG (*19*) and nominal pulsar distances (250pc and 288pc from (*14*)), for Geminga and PSR B0656+14, respectively). The diffusion coefficient D is obtained from the measured diffusion angle $\theta_0$ using Eq. S6, S7 and S9.

**Positron flux estimation**: We have performed electron diffusion calculations using the EDGE code, which is dependent on the *GAMERA* package (*41*).

The code assumes that pulsars emit as dipoles**;** therefore, their energy loss rate $L$ as a function of time $t$ since the pulsar was born is given by $L(t) = L_0\ (1+t/\tau)^{-2}$, where $\tau$ is the characteristic initial spin-down timescale, chosen to be 12000 yr (*28*) for both pulsars. From the previous equation the initial spin-down luminosity $L_0$ can be obtained for both pulsars: $L_0^{Gem}$=2.78 × $10^{37}$ erg $s^{-1}$ and $L_0^{PSR}$=4.0 × $10^{36}$ erg $s^{-1}$.

This result is used to calculate the energy density of electrons ((*21*), their equation 9) which is computed for every energy and every point in space at a time $t = t_{age}$, the age of the pulsar. Electrons lose energy while diffusing from synchrotron, inverse Compton, and bremsstrahlung losses.

Choosing $r = d_{Earth}$, the distance to Earth, the total flux in electrons and positrons at the Earth is $J(E)=c/4\pi\ f(r_{dist},t_{age},E)$. Electrons and positrons are considered to be produced in equal proportions, so the positron flux is just half of this flux, $J(E)/2$.

Using HAWC gamma-ray spectral information shown in Fig. S2 with the diffusion model fit, we obtain the normalization and spectral index of the electron injection spectrum, which is found to be harder than the gamma-ray spectra by ~0.1. Using the HAWC gamma-ray morphology, we get the diffusion coefficient. The rest of the parameters necessary to compute the energy density are shown in Table 1. The code computes the total gamma-ray spectrum produced by the source for the morphology measured.

Note that the gamma-ray spectrum derived from the electron diffusion code depends on the spectral index of injected electrons, the cooling of the electrons and the size of the source. The reason for this to be source-size dependent is that when integrating over small angles, we include more high energy electrons that cool down and do not reach larger distances because their diffusion radius is smaller. The size of the source is given by the fit of the diffusion model to the VHE gamma-ray profile of the source and the cooling of the electrons, including Klein-Nishina losses, is given by the magnetic field and photon fields assumed. The only free parameter is the spectral index of the electrons, that can be empirically extracted by running the code for different spectral indices and finding the one that best fits the data.

In order to include the statistical uncertainties in the positron flux, the parameters fitted from the HAWC data (flux normalization, spectral index, and diffusion angle) were sampled with a Gaussian distribution according to the statistical errors derived from the likelihood fit, with the constrain that the energy into electrons could not be larger than the energy injected by the pulsar.

**Systematic uncertainties**: Several detector systematic effects that can potentially affect the HAWC data are described in (*11*). The overall impact of these systematics is expected to be ±50% for the flux normalization, about ±0.2 for the spectral index, less than ±0.2 for $\log_{10}(E)$, and less than ±20% in the diffusion coefficient. The contribution of these effects combined is reflected in the dashed black line of Fig.3.

Apart from the previous systematic studies, other uncertainties coming from the choice in the parameters used have been also studied independently. Different diffusion coefficient spectral index $\delta$ values were tested based on the uncertainties reported by AMS. Moreover, the characteristic initial spin-down timescale $\tau$ was increased/decreased by a factor 3. In both cases the estimated positron flux remains well below the previously reported experimental results shown in Fig.3.

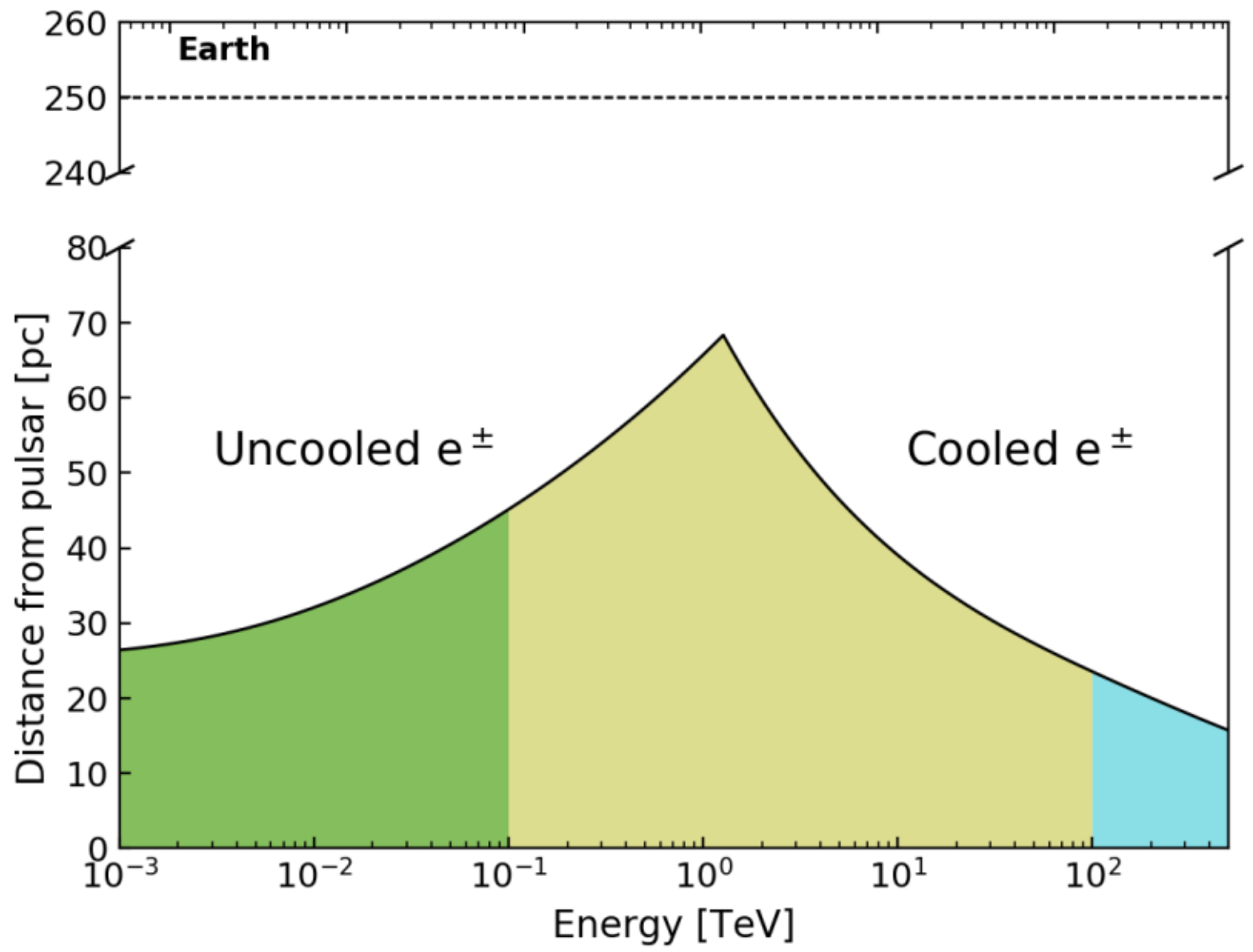

Fig. S1: Diffusion radius of electrons as a function of energy. Colors represent the regions shown in Fig. 1 (100 GeV in green, 1 TeV in light green and 100 TeV in blue). The plot shows that the low energy electrons do not diffuse very far, and the high energy ones do not live long enough. As a result, the electrons of around 1 TeV are the ones that travel further. This causes the sharply peaked profile of Fig. 3.

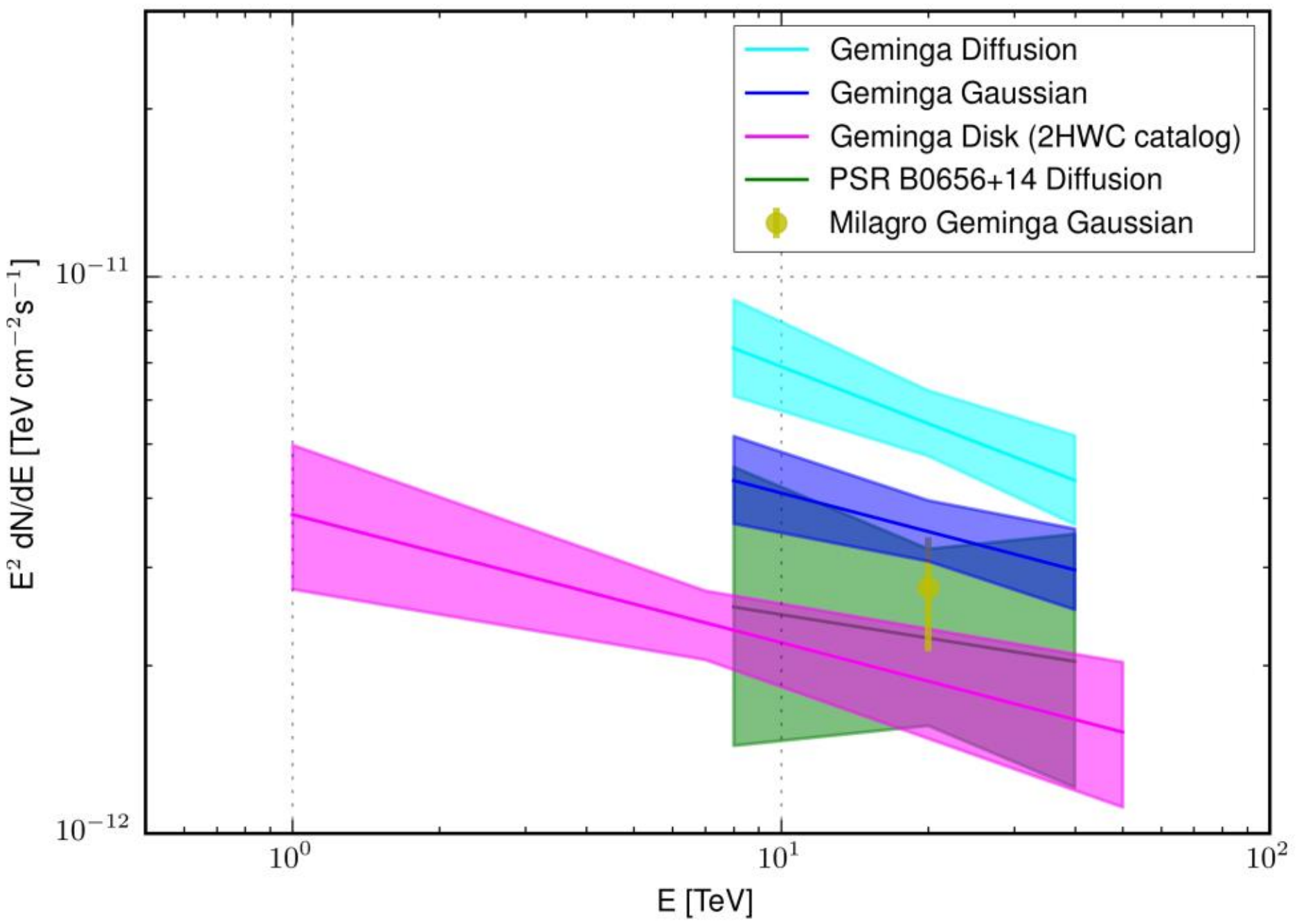

Fig. S2: **HAWC's gamma**-ray spectra of Geminga and PSR B0656+14. The blue, cyan and magenta lines correspond to the Geminga fluxes for different morphological assumptions. The apparent discrepancy between measurements is due to the different integrating regions considered for each morphology. The green line corresponds to the flux reported for PSR B0656+14 using a diffusion morphology. In all cases, the shaded band represent the statistical uncertainties of the fit. The point comes from the Geminga flux measured by Milagro in (*13*) for a Gaussian morphology.